\documentclass[submission,copyright,creativecommons]{eptcs}
\providecommand{\event}{GCM 2024}
\usepackage{graphicx} 
\usepackage{subcaption}
\usepackage{amssymb}
\usepackage{amsmath}
\usepackage{amsfonts}

\usepackage{iftex}
\ifpdf
\usepackage{underscore}         % Only needed if you use pdflatex.
\usepackage[T1]{fontenc}        % Recommended with pdflatex
\else
\usepackage{breakurl}           % Not needed if you use pdflatex only.
\fi

\usepackage{xspace}
\newcommand{\ie}{\emph{i.e.}\@\xspace}
\newcommand{\eg}{\emph{e.g.}\@\xspace}
\makeatletter
\newcommand*{\etc}{\@ifnextchar{.}{etc}{etc.\@\xspace}}
\makeatother

\usepackage{todonotes}

\usepackage[title]{appendix}

\newcommand{\ctrl}[1]{\textsf{#1}}
\newcommand{\rr}[1]{\mathtt{#1}}
\newcommand{\maxT}[1]{t_{\text{max}}}

\usepackage{hyperref}

\usepackage[capitalise]{cleveref}
\usepackage{orcidlink}
\usepackage{tikz}
\usepackage{tikz-qtree}
\usetikzlibrary{trees}
\usetikzlibrary{fit}
\usetikzlibrary{shapes}
\usetikzlibrary{shapes.geometric}
\usetikzlibrary{shapes.multipart}
\usetikzlibrary{shapes.gates.logic.US}
\usetikzlibrary{arrows}
\usetikzlibrary{arrows.meta}
\usetikzlibrary{calc,tikzmark}
\usetikzlibrary{positioning}

\tikzset{
  big edge/.style={
    green,
    thick,
  },
  big edgep/.style={
    big edge,
    -{Circle[fill=black,black,width=2,length=2,sep=-1]}
  },
  big pedge/.style={
    big edge,
    {Circle[fill=black,black,width=2,length=2,sep=-1]}-
  },
  big pedgep/.style={
    big edge,
    {Circle[fill=black,black,width=2,length=2,sep=-1]}-{Circle[fill=black,black,width=2,length=2,sep=-1]}
  },
  big edgec/.style={
    big edge,
    -{Bar[fill=green,green,width=4,length=0,sep=0]}
  },
  big pedgec/.style={
    big edge,
    {Circle[fill=black,black,width=2,length=2,sep=-1]}-{Bar[fill=black,black,width=4,length=0,sep=0]}
  },
  big region/.style={
    draw,
    rectangle,
    rounded corners=1.5,
    dashed,
    dash pattern=on 1pt off 1pt,
    thin,
    gray,
  },
  big site/.style={
    big region,
    fill=gray!60,
    text=black,
  },
  big react/.style={
    black,
    thick,
    -stealth,
    line width=3,
    shorten <=3,
    shorten >=3,
  },
  big react rev/.style={
    black,
    thick,
    stealth-stealth,
    line width=3,
    shorten <=3,
    shorten >=3,
  },
  big inst map/.style={
    thick,
    -stealth,
    blue,
    dashed
  },
  lbl/.style={
    font=\tiny\sf,
    inner sep=1,
  },
  lbl conc/.style={
    font=\tiny,
    inner sep=1,
  }
}

\DeclareMathOperator{\rrul}{\mathrel{\frac{\raisebox{0.75mm}{\begin{scriptsize}\ensuremath{\hspace*{1mm}\ \hspace*{1mm}}\end{scriptsize}}}{}} \joinrel{\!\!\blacktriangleright}}

\crefname{lstlisting}{Listing}{Listings}
\Crefname{lstlisting}{Listing}{Listings}

\usepackage{listings}

\definecolor{solarized@base03}{HTML}{002B36}
\definecolor{solarized@base02}{HTML}{073642}
\definecolor{solarized@base01}{HTML}{586e75}
\definecolor{solarized@base00}{HTML}{657b83}
\definecolor{solarized@base0}{HTML}{839496}
\definecolor{solarized@base1}{HTML}{93a1a1}
\definecolor{solarized@base2}{HTML}{EEE8D5}
\definecolor{solarized@base3}{HTML}{FDF6E3}
\definecolor{solarized@yellow}{HTML}{B58900}
\definecolor{solarized@orange}{HTML}{CB4B16}
\definecolor{solarized@red}{HTML}{DC322F}
\definecolor{solarized@magenta}{HTML}{D33682}
\definecolor{solarized@violet}{HTML}{6C71C4}
\definecolor{solarized@blue}{HTML}{268BD2}
\definecolor{solarized@cyan}{HTML}{2AA198}
\definecolor{solarized@green}{HTML}{859900}

\lstdefinelanguage{bigrapher}{
    keywords={fun, brs, end, sbrs, pbrs, abrs, begin, init, atomic, preds,
      rules, actions, big, ctrl, float, int, react, share, by},
    morekeywords={in, if, param, ctx},
    numberstyle=\tiny\color{solarized@base01},
    keywordstyle=\color{solarized@green},
    stringstyle=\color{solarized@cyan}\ttfamily,
    commentstyle=\color{solarized@base01},
    emphstyle=\color{solarized@red},
    comment=[l]{\#},
}

\lstnewenvironment{bigrapher}[1][] {
  \lstset{
    basicstyle=\scriptsize\ttfamily,
    mathescape=true,
    breaklines=true,
    sensitive=true,
    numbers=left,
    numberstyle=\tiny,
    numbersep=3pt,
    xleftmargin=8pt,
    escapeinside={:}{:},
    showstringspaces=false,
    flexiblecolumns=false,
    frame=tb,
    language=bigrapher,
    #1
  }} {}

\title{Modelling Real-time Systems with Bigraphs
\author{Maram Albalwe\orcidlink{0009-0007-4785-9479}
\institute{University of Glasgow \\ Glasgow, UK}
\institute{University of Tabuk \\ Tabuk, Saudi Arabia}
\email{m.albalwe.1@research.gla.ac.uk}
\and
Blair Archibald\orcidlink{0000-0003-3699-6658}\qquad Michele Sevegnani\orcidlink{0000-0001-6773-9481}
\institute{University of Glasgow\\
 Glasgow, UK}
\email{\{blair.archibald, michele.sevegnani\}@glasgow.ac.uk}
}
}
\def\authorrunning{Maram Albalwe, Blair Archibald \& Michele Sevegnani}
\def\titlerunning{Modelling Real-time Systems with Bigraphs}

\begin{document}
\maketitle 

\begin{abstract}
Bigraphical Reactive Systems (BRSs) are a graph-rewriting formalism describing systems evolving in two dimensions: spatially, \eg a person in a room, and non-spatially, \eg mobile phones communicating regardless of location. Despite use in domains including communication protocols, agent programming, biology, and security, there is no support for real-time systems. We extend BRSs to support real-time systems with a modelling approach that uses multiple perspectives to represent digital clocks. We use Action BRSs, a recent extension of BRSs, where the resulting transition system is a Markov Decision Process (MDP). This allows a natural representation of the choices in each system state: to either allow time to pass or perform a specific action. We implement our proposed approach using the BigraphER toolkit, and demonstrate the effectiveness through multiple examples including modelling cloud system requests.
\end{abstract}

\section{Introduction}
Bigraphs are a computational model where systems are described based on two types of relationships: spatially using \emph{nesting} (and parallel adjacency) and non-local linking through hyperlinks regardless of locations.
Like standard graphs, bigraphs have an equivalent diagrammatic and algebraic representation.
Bigraphical Reactive Systems (BRSs) equip bigraphs with a set of reaction rules that specify how a system evolves over time, \ie reaction rules substitute sub-bigraphs with other bigraphs.

The standard theory of bigraphs has been extended to model a wider range of systems \eg stochastic bigraphs~\cite{stochastic} assign rates to reaction rules, bigraphs with sharing~\cite{sharing} allow intersecting locations, directed bigraphs~\cite{directed} associate directions to links, conditional bigraphs~\cite{condition} add conditions to rules, and probabilistic and action bigraphs~\cite{Probablistic} support probabilistic and non-deterministic behaviour.
Using these extensions, bigraphs have been successfully used to model a variety systems including mixed-reality games~\cite{savannah}, cloud systems~\cite{cloudBig}, self-adaptive fog systems~\cite{fogBig}, sensor
network infrastructure~\cite{sensorBig}, and rational agents~\cite{BDIAgent}.

An extension that has not been explored is real-time bigraphs that can model systems exhibiting non-deterministic behaviour that is controlled by time constraints.
While the non-deterministic aspects of real-time systems could be modelled by action bigraphical reactive systems (ABRSs) in which the underlying Markov Decision Process (MDP) semantics allows for a transition choice at each state, there is currently no notion of \emph{clock constraints} in the theory, \eg specifying that after $3$ time units have passed, a given action \emph{must} occur. 

We propose a modelling strategy to encode timed systems within ABRSs through the well-known MDP-based digital clock approximation. 
We introduce a clock, as a new bigraph entity, for each timed entity in the system and we place these in a separate region so all the clocks  can be manipulated without polluting the actual system model.
This allows one reaction rule to advance all clocks at once, and this reduces the state space overhead of the approach.
We mirror the real-time passage by introducing a global clock that controls the system execution time, \ie system terminates when we reach the time limit of this clock, and we explicitly model the non-deterministic behaviour, \ie at each state there is a choice between taking an action or allowing time to pass when the constraints are not yet satisfied.

We make the following research contributions:
\begin{itemize}
    \item We propose a modelling strategy to express clocks within action bigraphs.
    
    \item We define a set of reaction rules to model clock constraints in real-time systems through a digital clocks approximation.

    \item We illustrate how to use our approach in practice by partially modelling the timed aspects of a cloud computing scenario.

    \item We implement this approach in BigraphER~\cite{bigrapher} to show this is not just a theoretical contribution, but a practical one.
\end{itemize}
\paragraph{\textbf{Outline.}} 
We give an informal description of bigraphs and BRSs in \Cref{background}. \Cref{timedbigraphs} provides a description of non-deterministic models and shows how to formalise digital clocks within ABRSs. We illustrate our approach by providing a BRS implementation for a simple scenario modelled as a (probabilistic) timed automaton and a model of cloud system requests in \Cref{examples}. \Cref{relatedwork} gives a brief literature review, and we conclude in \Cref{conclusion}. 

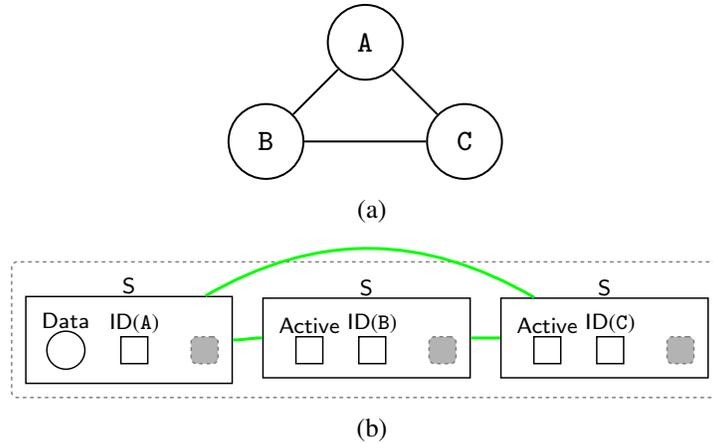
\begin{figure}
\centering
\begin{subfigure}[b]{0.6\textwidth}
\centering
\resizebox{0.4\textwidth}{!}{\begin{tikzpicture}[
    box/.append style = {draw, circle, align=center, font=\tiny},
    arr/.append style = {draw, -latex},
]

\node[box] (location1) {$\mathtt{A}$};

\node[box, below left of= location1] (location2) {$\mathtt{B}$};

\node[box, below right of= location1] (location3) {$\mathtt{C}$};

\path (location1) edge  (location2);
\path (location1) edge  (location3);
\path  (location3) edge  (location2);
\end{tikzpicture}}
\caption{}\label{exampleA}
\end{subfigure}

\begin{subfigure}[b]{0.6\textwidth}
\resizebox{\textwidth}{!}{\begin{tikzpicture}[
  ,
idle/.append style = {draw},
data/.append style = {draw, circle},
c/.append style = {draw},
b/.append style = {draw},
a/.append style = {draw},
iD/.append style = {draw}
  ]
    
\node[data,  label={[inner sep=0.5, name=n1l]north:{\tiny {\sf Data}}}] (n1) {};
\node[idle, right=2 of n1, label={[inner sep=0.5, name=n3l]north:{\sf\tiny Active}}] (n3) {};
\node[idle, right=2 of n3, label={[inner sep=0.5, name=n5l]north:{\sf\tiny Active}}] (n5) {};
\node[big site, right=1 of n1,] (s0){};
\node[big site, right=1 of n3,] (s1){};
\node[big site, right=1 of n5,] (s2){};

 \node[iD, left=.40 of s0, label={[inner sep=0.5, name=n30l]north:{\tiny {\sf ID}(\texttt{A})}}] (n30) {};

 \node[iD, left=0.40 of s1, label={[inner sep=0.5, name=n31l]north:{\tiny {\sf ID}(\texttt{B})}}] (n31) {};

 \node[iD, left=0.40 of s2, label={[inner sep=0.5, name=n32l]north:{\tiny {\sf ID}(\texttt{C})}}] (n32) {};

\node[a, fit=(n1)(n1l)(s0)(n30)(n30l), label={[inner sep=0.5, name=n0l]north:{\sf\tiny S}}] (n0) {};
\node[b, fit=(n3)(n3l)(s1)(n31)(n31l), label={[inner sep=0.5, name=n2l]north:{\sf\tiny S}}] (n2) {};
\node[c, fit=(n5)(n5l)(s2)(n32)(n32l), label={[inner sep=0.5, name=n4l]north:{\sf\tiny S}}] (n4) {};

\node[big region, fit=(n4)(n4l)(n2)(n2l)(n0)(n0l)] (r0) {};

\draw[big edge] (n0) to[out=0,in=180] (n2);
\draw[big edge] (n0) to[out=30,in=150] (n4);
\draw[big edge] (n2) to[out=0,in=180] (n4);

\end{tikzpicture}}
\caption{}\label{exampleB}
\end{subfigure}
\caption{(a) Example network topology with three
sensors; (b) corresponding bigraph with data to be transmitted by sensor \texttt{A} and the status of each sensor (\eg Active).}
\label{fig:example}

\end{figure}

\section{Background}\label{background} 

\subsection{Bigraphs}
Introduced by Milner~\cite{milner}, Bigraphs provide a powerful diagrammatic representation for modelling systems that evolve in both spatial and non-spatial dimensions.
Bigraphs represent the structure of a system through two relations over its entities: a \emph{place graph} that encodes their spatial arrangement, \eg a person in a room, and a \emph{link graph} that specifies, through hyper-edges, non-spatial relations, \eg communication capabilities between network devices.
We give an informal description of bigraphs using the example in \Cref{fig:example}.
A comprehensive formal description can be found elsewhere~\cite{milner}. 
Bigraphs have both a diagrammatic representation and \emph{equivalent} algebraic notation. 
We use the diagrammatic notion throughout this paper.

\Cref{exampleA} shows a simple network topology consisting of three sensors connected to each other by communication links. Data may be exchanged between two sensors when the receiver is active. A corresponding bigraph representation is in \Cref{exampleB}. Sensors are represented as \emph{entities} of type $\ctrl{S}$ that have \emph{nested} identifiers (also just a different type of entity) $\ctrl{ID}(\mathtt{A})$, $\ctrl{ID}(\mathtt{B})$, and $\ctrl{ID}(\mathtt{C})$. 
The sender contains an entity $\ctrl{Data}$ that can be sent to other sensors when they are in the $\ctrl{Active}$ mode.
We sometimes draw different entity types by using different shapes or colours \eg we have used squares for identifiers and a circles for $\ctrl{Data}$.
Entities can be \emph{atomic}, \eg $\ctrl{Data}$, meaning they have no children, or contain any number of other entities. 
We allow entities to be parameterised to represent families of entities, \ie $\ctrl{ID}(x)$ specifies a new entity for every possible value of $x$ (which may be string, integer, or float typed).
Sites---filled dashed rectangles---represent unspecified bigraphs: an arbitrary bigraph, including the empty bigraph, might exist there.
Bigraphs may consist of more than one region---clear dashed rectangles---which represent adjacent parts of the system.
For the link graph, entities have a fixed \emph{arity} that represents the number of \emph{links} they must have (the green edges). Links are hyperlinks that may connect $1$-to-$n$. Open links---a link that has a name---
indicates a link that may connect elsewhere, \ie to currently unspecified entities. Links may also be closed, a $1$-to-$0$ hyperedge, which is shown by an orthogonal line at the end of the link. 
Open names, sites, and regions allow model composition: regions of one bigraph can be placed in the sites of another and like-names joined. 
This forms the basis of the rewriting theory.

\subsection{Bigraphical Reactive Systems (BRSs)}\label{BRS}
\begin{figure}
	\centering
%	\resizebox{0.7\textwidth}{!}{\input{tikztext/simpleExampleRule.tex}}
	\resizebox{0.7\textwidth}{!}{\includegraphics[width=\linewidth]{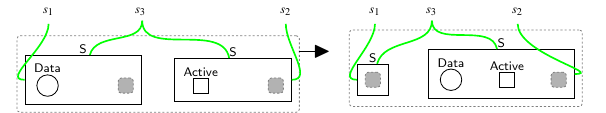}}
	\caption{Reaction rule $\rr{send\_Data}$ applies when the receiver is active.}
	\label{ExampleRule}
\end{figure}

A bigraph represents a system at a single point in time. To model dynamic behaviour, Bigraphical Reactive Systems (BRSs) equip bigraphs with a set of \emph{reaction rules} specifying how the system may evolve.
Reaction rules take the form $L \rrul R$ meaning that when the rule is applicable to a state $S$ (state $S$ \emph{matches} bigraph $L$) the system evolves by replacing an occurrence of bigraph $L$ in $S$ with bigraph $R$. BRSs form a transition system that has an initial state (\emph{bigraph}), and the transition from one state (\emph{bigraph}) to another is defined by generating all possible rewrites (\emph{reactions}). For example, the rule in \Cref{ExampleRule} is applicable where there is a sensor that has \ctrl{Data} to send to another \ctrl{Active} sensor in its range. 

We control the execution of a set of reaction rules via \emph{priority classes}.
That is $\{\texttt{r}_1, \texttt{r}_2\} < \{\texttt{r}_3\}$ means rules $\texttt{r}_1$ and $\texttt{r}_2$ can be applied only if it is not possible to apply $\texttt{r}_3$.
Similarly to entities, rules can be parameterised to allow multiple rule applications over a predefined set of values. 

\section{Modelling Time with Action Bigraphs}\label{timedbigraphs}
The big idea is that by utilising Action Bigraphs~\cite{Probablistic} we can encode timed systems.
This is based on a well-known approximation of (probabilistic) timed automata to Markov Decision Processes~\cite{digitalClocks}.
Intuitively, we extend the usual action semantics to allow two forms of action: \emph{discrete actions} that encode system events (which may only be enabled at some time), and \emph{time} actions that progress time.

Action Bigraphical Reactive Systems (ABRSs) allow a choice of probability distributions at each rewriting step by assigning \emph{weights} to the reaction rules, and by giving action labels to specific sets of reaction rules.
An action is enabled/possible whenever there is a reaction rule from the set which is enabled.
The resulting transition system is a Markov Decision Processes (MDP)~\cite{MDP1} which can be verified using off-the-shelf model checkers such as PRISM~\cite{prism}. An MDP is a tuple $(S, s_0, A, Step)$ where $S$ is a set of states with a predefined initial state $s_0 \in S$. For each state, we can choose an (enabled) action $a \in A$ which gives an associated probability distribution over future states as specified by $Step$. 

\begin{figure}[!htb]
\centering
\begin{subfigure}[b]{0.75\textwidth}
\centering
%\resizebox{\textwidth}{!}{\input{tikztext/sensorAsendaction.tex}}
\resizebox{\textwidth}{!}{\includegraphics[width=\linewidth]{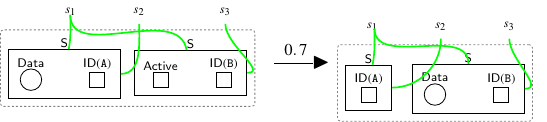}}
\caption{Sensor \texttt{A} sends to sensor \texttt{B} with weight 0.7}\label{Asend}
\end{subfigure}
\begin{subfigure}[b]{0.75\textwidth}
%\resizebox{\textwidth}{!}{\input{tikztext/sensorAsendToB.tex}}
\resizebox{\textwidth}{!}{\includegraphics[width=\linewidth]{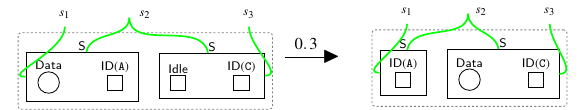}}
\caption{Sensor \texttt{A} sends to sensor \texttt{C} with weight 0.3}\label{Breceives}
\end{subfigure}
\begin{subfigure}[b]{0.65\textwidth}
    \centering
%\resizebox{\textwidth}{!}{\input{tikztext/ActionBigSensorsSend.tex}}
\resizebox{\textwidth}{!}{\includegraphics[width=\linewidth]{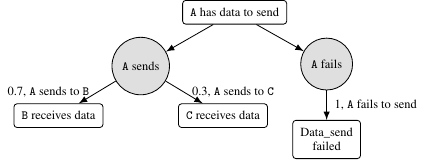}}
\caption{MDP generated by the ABRS.}\label{actBig}
\end{subfigure}
\caption{Example Action Bigraphical Reactive System: probabilistic reaction rules using weights.
}
\label{fig:actionBigexample}
\end{figure}

To show how ABRSs model non-deterministic behaviour of real-time systems where underlying transition system is an MDP, 
we use the bigraph example shown in \Cref{exampleB}.
That is for sensor $\mathtt{A}$, data can either be sent successfully to other sensors or fail to send.
We also can permit sensor $\mathtt{A}$ to send \ctrl{Data} with a bias by replacing the $\rr{send\_Data}$ rule (\Cref{ExampleRule}) with the two weighted (as shown in \Cref{fig:actionBigexample}).
For example, given $weight\,=\,0.7$ to $\mathtt{B}$ and $weight\,=\,0.3$ to $\mathtt{C}$, then sending to $\mathtt{B}$ is more than twice as likely. 
We use \emph{bigraphs} throughout this paper to mean \emph{Action Bigraphical Reactive Systems (ABRSs)}. Using BigraphER~\cite{bigrapher}, an open-source toolkit for working with bigraphs (and the supporting ABRSs), we can export the corresponding MDPs transition system to be formally checked. 
Importantly, action bigraphs still respect any rule priorities.
This means an action will fire with any high priority rule before firing with those of lower priorities.
We use this feature in our encoding of clock invariants.

While MDPs support uncertainty modelling for dynamic environments, Probabilistic Timed Automata (PTAs)~\cite{timedAutomata} extend this by incorporating timing constraints which offer a more comprehensive framework for modelling and reasoning about real-time systems. 
A PTA is defined as a tuple in the form $(L, l_0, E, A, C, I)$ where $L$ is a set of locations (\ie states) with $l_0 \in L$ a start location.
$E$ is a set of edges that represents the possible transitions between locations while $A$ is a set of (discrete) actions that may occur in the system.
$C$ is a finite set of clocks and $I$ is a set of clock invariants associated to each location.
While these invariants could be flexible, practically they are typically used to force a transition from a location, \eg an invariant $x \le 2$ forces a location move at time $2$ if we have not already left the location (alongside transition conditions these usually encode a ``move within a maximum of 2 time units'' semantics).
Transitions are associated to actions and probability distributions, with the addition of \emph{clock constraints} (\eg $x > 3$) and \emph{clock resets} (\eg $x := 0$).

In the \emph{digital clocks approximation} we give the semantics of PTAs as a Markov Decision Process where states (including initial) are all the locations where clock invariants are met, and  we form a single action set consisting of both actions manipulating time and user-specified discrete actions as follows:
\begin{itemize}
    \item a \emph{discrete} action is enabled if all clock constraints (given in $E$) associated with the transition are satisfied;
    \item a \emph{time} action is available while invariants associated to $l \in L$ are satisfied as time elapses. That is, we cannot progress time if something must happen beforehand.
\end{itemize}

Since we use action bigraphs, we can associate discrete transitions with a probability distribution. If required we can, for example, draw with uniform probability to recover semantics for a non-probabilistic timed automata.

\subsection{Digital Clocks in Bigraphs}\label{bigClocks}
Standard BRSs evolve to a new state whenever there is a match of a reaction rule\footnote{Subject to any requirements about rule priorities and conditions.}.
In this sense they are non-deterministic but not explicitly so, \ie we cannot label particular actions without ABRSs.
Models of real-time systems need to both find enabled matches, and meet any requirements introduced by clock constraints. 
We propose a modelling technique that encodes digital clocks as \emph{entities} to model real-time systems. 
Although PTAs can work with real-valued clocks, for the digital clocks approximation we fix clock values to non-negative integers and bound the total runtime of the system (to ensure a finite action set which means the models can be analysed through existing MDP reachability techniques).
As for the standard digital clocks approximation, our approach does not support strict inequalities and comparisons between clocks. 
As we have action bigraphs, these clocks can live within the bigraph model itself, and a separate type of semantics is not needed (unlike for probabilistic bigraphs for example).
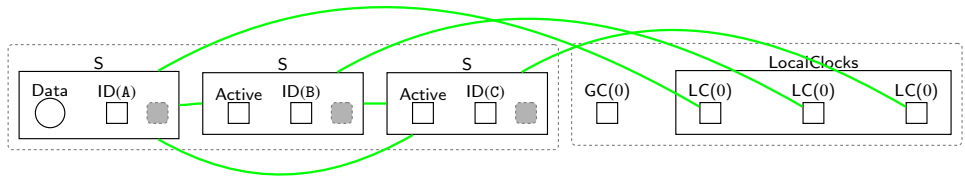
\begin{figure}
\centering
\resizebox{0.8\textwidth}{!}{\begin{tikzpicture}[
  ,
mins/.append style = {draw},
localClock/.append style = {draw},
lC/.append style = {draw},
idle/.append style = {draw},
globalClock/.append style = {draw},
gC/.append style = {draw},
data/.append style = {circle, draw},
iD/.append style = {draw},
c/.append style = {draw},
b/.append style = {draw},
a/.append style = {draw}
  ]
    
\node[data,  label={[inner sep=0.5, name=n1l]north:{\sf\tiny Data}}] (n1) {};
\node[idle, right=2.00 of n1, label={[inner sep=0.5, name=n3l]north:{\sf\tiny Active}}] (n3) {};
\node[idle, right=2.00 of n3, label={[inner sep=0.5, name=n5l]north:{\sf\tiny Active}}] (n5) {};
\node[gC, right=2.00 of n5, label={[inner sep=0.5, name=n8l]north:{\sf\tiny GC($0$)}}] (n8) {};
\node[big site, right=1.00 of n1,] (s0){};
\node[big site, right=1.00 of n3,] (s1){};
\node[big site, right=1.00 of n5,] (s2){};

\node[lC, right=1.00 of n8, label={[inner sep=0.5, name=n11l]north:{\sf\tiny LC($0$)}}] (n11) {};
\node[lC, right=1.00 of n11, label={[inner sep=0.5, name=n13l]north:{\sf\tiny LC($0$)}}] (n13) {};
\node[lC, right=1.00 of n13, label={[inner sep=0.5, name=n15l]north:{\sf\tiny LC($0$)}}] (n15) {};
\node[a, fit=(n1)(n1l)(s0)(n30)(n30l), label={[inner sep=0.5, name=n0l]north:{\sf\tiny S}}] (n0) {};
\node[b, fit=(n3)(n3l)(s1)(n31)(n31l), label={[inner sep=0.5, name=n2l]north:{\sf\tiny S}}] (n2) {};
\node[c, fit=(n5)(n5l)(s2)(n32)(n32l), label={[inner sep=0.5, name=n4l]north:{\sf\tiny S}}] (n4) {};
\node[iD, right=0.50 of n1, label={[inner sep=0.5, name=n30l]north:\tiny {\sf ID}(\texttt{A})}] (n30) {};
\node[iD, right=0.50 of n3, label={[inner sep=0.5, name=n31l]north:\tiny {\sf ID}(\texttt{B})}] (n31) {};
\node[iD, right=0.50 of n5, label={[inner sep=0.5, name=n32l]north:\tiny {\sf ID}(\texttt{C})}] (n32) {};
\node[localClock, fit=(n11)(n11l)(n13)(n13l)(n15)(n15l), label={[inner sep=0.5, name=n9l]north:{\sf\tiny LocalClocks}}] (n9) {};
\node[big region, fit=(n4)(n4l)(n2)(n2l)(n0)(n0l)] (r0) {};

\node[big region, fit=(n9)(n9l)(n8)(n8l)] (r1) {};
\draw[big edge] (n0) to[out=0,in=180] (n2);
\draw[big edge] (n0) to[out=-30,in=-150] (n4);
\draw[big edge] (n0) to[out=30,in=150] (n11);
\draw[big edge] (n2) to[out=0,in=180] (n4);
\draw[big edge] (n2) to[out=30,in=150] (n13);
\draw[big edge] (n4) to[out=30,in=150] (n15);

\end{tikzpicture}}
\caption{Example from \Cref{exampleB} extended with clocks perspective.
}\label{rule:clockPerspictive}
\end{figure}
We show our approach by an example, and the main idea is in \Cref{rule:clockPerspictive}.
We put all clocks into their own parallel region which we call the \emph{clocks perspective}.
A global clock is represented by an entity family $\mathsf{GC}(n)$---\ie there is one entity for each $0 \leq n \leq \maxT{}$ for some max time $\maxT{}$---and is used to model \emph{wall clock time}.
Wall-time cannot be reset. 
Local clocks are entity families $\mathsf{LC}(n) 
\: \text{where}\: 0\, \le n \, \le \, \maxT{}$ for each timed entity. 
For clarity of modelling, we place these in $\ctrl{LocalClocks}$, although this is not strictly required.
We identify a set of \emph{timed} entities and expand their arity by 1 to link them to a specific local clock.
Using this, we can \emph{identify} a clock through the linked entity, \ie clocks do not need specific identifiers of their own.
Not all entities in the system will own a clock, \eg $\ctrl{Data}$ is not timed.
We assume all clocks are initialised to $0$ in the initial state.

The approach to place clocks within their own region is a design choice, and because of the flexibility of bigraphs, other choices would be possible.
For example, we could \emph{nest} local clocks within particular timed entities instead of linking to them.
We chose the extra region as it does not clutter any existing model, \ie we can convert an existing model to a timed representation without significant changes (other than arity) in existing regions.

\begin{figure}
\centering
%\resizebox{0.8\textwidth}{!}{\input{tikztext/clockadvance.tex}}
\resizebox{0.8\textwidth}{!}{\includegraphics[width=\linewidth]{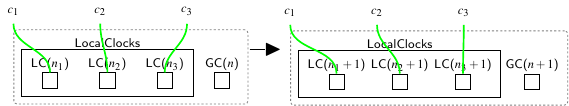}}
\caption{Reaction rule $\rr{clock\_advance}(n_1, n_2, n_3, n)$ for a system with three timed entities.}\label{rule:clockAdvance}
\end{figure}

\subsection{Reaction Rules for Digital Clocks}
In our digital clocks approximation we have two main types of action: \emph{discrete} actions which are user-defined and system specific, and \emph{time} actions that deal with the passage of time.
As actions in ABRSs are modelled as sets of reaction rules, the main challenge here is defining appropriate reactions for each type of action.

For timing actions, the main rule is \texttt{clock\_advance}$(n_1, n_2, \dots, n)$ shown in \Cref{rule:clockAdvance} (for a system with three timed entities). 
This rule advances all local and global clocks simultaneously. 
All clocks advance at the same speed (one unit per application).
This is a parameterised rule, which, like paramterised entities, defines a family of rules \ie one for each possible value of the parameters.
The $\mathit{tick}$ action consists of the set of all possible $\rr{clock\_advance}(n_1,n_2,\dots, n)$ rules for parameters drawn from $n, n_1, n_2, \ldots \leq \maxT{}$, for some max time $\maxT{}$.
Due to the parameters only one instance of $\rr{clock\_advance}$ will ever be enabled meaning this action always advances time with probability one.

For \emph{discrete} actions we extend existing system rules whenever a time constraint must be met.
We take rules that have a standard (untimed) definition of a bigraphs rule ($L \rrul R$) and, like with the clocks perspective, utilise parallel regions to link timed entities with their clocks as follows:
\begin{gather*}
/c\, (\widehat{L_{c}} \,\parallel\, \mathsf{LC}(n)_{c}) \rrul /c\, (\widehat{R_{c}} \,\parallel\, \mathsf{LC}(n')_{c})
\end{gather*}
where $n' \in \{n,0\}$, notation $\widehat{\_}$ indicates the transformation over bigraphs described in \cref{bigClocks} in which the entity type of timed entities has its arity increased by one, \ie each entity has one more name.
The subscript $c$ indicates the name used to link a specific timed entity to a new clock placed in a \emph{different} regions as specified by the~$\parallel$ operator.
Other entities are forbidden from linking over $c$ as the link is \emph{closed} (with $/c$). 
Importantly, this change also makes existing rules into parameterised rules (over parameter $n$).
This is used later to encode timing constraints, \eg we chose a set of parameters that define at what (local) \emph{time} a particular rule can be applied.
For rules that are already parameterised, we can simply add an additional parameter for the clock.

\begin{figure}
\centering
%\resizebox{0.8\textwidth}{!}{\input{tikztext/timedRule2.tex}}
	\resizebox{0.8\textwidth}{!}{\includegraphics[width=\linewidth]{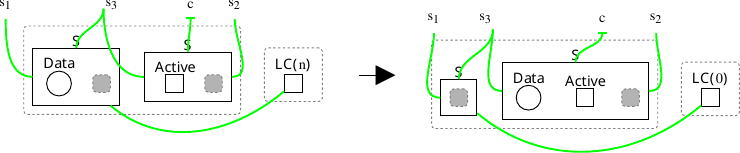}}
    \caption{Reaction rule(s) $\rr{sending\_data}(n)$. Time constraints are encoded by setting $2 \leq n \leq \maxT{}$.
   }\label{rule:sendingData}
\end{figure}

As an example we extend our rule in \Cref{ExampleRule}, that allows sensors to send data, to only fire when the receiver is active, \emph{and} at least two time units have passed \eg $\mathsf{LC}(n) \geq 2$. 
As we are modelling an inequality we cannot do this with a single reaction rule and instead we provide a new parameterised rule $\rr{sending\_data}(n)$, shown in \Cref{rule:sendingData}, that explicitly links the sending sensor to its local clock.
The rule is only valid for $n \in \{2, 3, \dots, \maxT{}\}$ so this is the family of rules we generate.
In this case, we include a clock reset on the right-hand-side of the rule, and clocks may only be reset during the application of a rule.

Note that the $\rr{clock\_advance}$ rule has the same priority as the corresponding rules whose applications are triggered by clock constraints. This allows time to pass until the clock valuation satisfies the first invariant. In a such state, the non-deterministic choices are applicable \ie the system can take a discrete action or permit the time to pass if it is not the last valid clock valuation in which an action must occur. 

We must also encode any location-based clock invariants, \eg locations that are only valid for $x \le 2$.
In practice this is used to force a transition to occur rather than allowing an indefinite wait within a particular location, and is almost always $true$ (allowing indefinite waits) or a $\le$ constraint.
To encode these invaraints we make use of priority classes in the ABRS.
For a state invariant, \eg $t \le x \le n$, we add any outgoing transition rules $\rr{r}(n)$ such that $\{\rr{clock\_advance}(\dots), \rr{r}(t), \rr{r}(t+1), \dots, \rr{r(n-1)}\}\} < \{ \rr{r}(n)\}$.
This means $\rr{r}(n)$ applies \emph{before} any clock updates and \emph{forces} a state transition before the invariant is broken.

\section{Examples and Implementation}\label{examples}
We implement our approach in BigraphER which also generates the MDPs (that are the semantics for our digital clocks representation). The exported MDPs can be formally checked using standard (probabilistic) model checkers, \eg PRISM, allowing us to benefit from existing model checking algorithms.
 
\begin{figure}
\centering
%\resizebox{0.5\textwidth}{!}{\input{tikztext/PTAClockExample.tex}}
\resizebox{0.5\textwidth}{!}{\includegraphics[width=\linewidth]{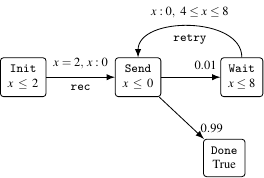}}
\caption{A probabilistic timed automata example model for a simple sending data process from~\cite{PTAdigitalClock}.
}
\label{PTAExample}
\end{figure}

Modelling a timed system with our approach follows these general steps:
\begin{itemize}
    \item Define sets of clock valuations/invariants according to the system requirements.
    \item Define a new clock perspective and add clock entities equal to the number of the timed entities required, and one global clock. Increase the arity of timed entities by 1, and link to the local clock. 
    All clocks are initialised to 0.
    \item For a given $\maxT{}$ and number of timed entities, generate the set of rules in the form
    \[
    \rr{clock\_advance}(\{0, \dots, \maxT{}\},\{0, \dots, \maxT{}\}, \dots)
    \]
    \item Extend system rules by adding in the clock perspective when required to restrict their application based on the time constraints. Parameters for these rules must meet the clock invariants. Any state-based invariants are encoded using priorities.
\end{itemize}

We illustrate our approach by encoding two different examples: a probabilistic timed automata example, and requests allocation in a cloud system~\cite{timedBig}. 
The BigraphER models for both examples are in the appendix.

\begin{figure}[!htb]
\centering
\begin{subfigure}[b]{0.6\textwidth}
\centering
%\resizebox{0.7\textwidth}{!}{\input{tikztext/inittransition.tex}}
\resizebox{0.7\textwidth}{!}{\includegraphics[width=\linewidth]{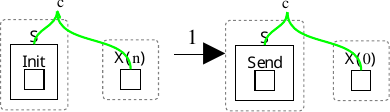}}
\caption{Reaction rule $\rr{init\_transition}(n)$, $n \in\{0,1,2\}$}\label{initTran}
\end{subfigure}

\begin{subfigure}[b]{0.6\linewidth}
\centering
%\resizebox{0.7\textwidth}{!}{\input{tikztext/sendtransitionsuccess.tex}}
\resizebox{0.7\textwidth}{!}{\includegraphics[width=\linewidth]{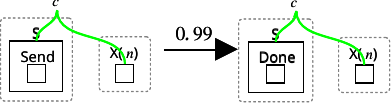}}
\caption{Reaction rule $\rr{send\_transition\_success}(n)$, $n=0$}\label{sendTranSucc}
\end{subfigure}

\begin{subfigure}[b]{0.6\linewidth}
\centering
%\resizebox{0.7\textwidth}{!}{\input{tikztext/sendtransitionfail.tex}}
\resizebox{0.7\textwidth}{!}{\includegraphics[width=\linewidth]{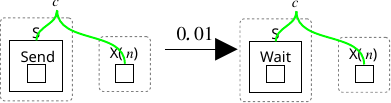}}
\caption{Reaction rule $\rr{send\_transition\_fail}(n)$, $n=0$}\label{sendTranFail}
\end{subfigure}

\begin{subfigure}[b]{0.6\linewidth}
\centering
%\resizebox{0.7\textwidth}{!}{\input{tikztext/waittransition.tex}}
\resizebox{0.7\textwidth}{!}{\includegraphics[width=\linewidth]{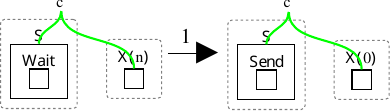}}
\caption{Reaction rule $\rr{wait\_transition}(n)$, $n \in \{4,5,6,7,8\}$}\label{waitTran}
\end{subfigure}

\caption{Reaction rules for the PTA example.}
\label{fig:PTAModelrules}
\end{figure}

\begin{figure}[!htb]
\centering
\resizebox{0.7\textwidth}{!}{\begin{tikzpicture}[
    box/.append style = {draw, ellipse, rounded corners=1.5, align=center, font=\tiny},
    act/.append style = {draw, ellipse, font=\tiny, fill=gray!25},
   % box/.append style = {draw, rectangle, align=center, font=\tiny},
    arr/.append style = {draw, -latex},
    arrr/.append style = {draw, dotted},
]

\node[box] (1) {clock\_X=0, Init state};
\node[act,below left=0.5 of 1] (2)  {tick};
\node[act,below right =0.7 of 1] (3)  {rec};
\node[box,below =0.5 of 2] (4)  {clock\_X=1, Init state};
\node[act,below left =0.5 of 4] (5)  {tick};
\node[act,below right =0.5 of 4] (6)  {rec};
\node[box,below =0.5 of 5] (7)  {clock\_X=2, Init state};
\node[act,below =0.5 of 7] (8)  {rec};

%\node[box,below =4.7 of 1] (9)  {X=0, Send state};
\node[box,below = 4 of 1, xshift=5] (9)  {clock\_X=0, Send state};

\node[act,below  =0.5 of 9] (10)  {send};
\node[box,below left =0.5 of 10] (11)  {clock\_X=0, Wait state};
\node[box,below right =0.5 of 10] (12)  {clock\_X=0, Done state};

\node[act,below =0.5 of 11] (13)  {tick};
\node[box,below =0.5 of 13] (14)  {clock\_X=1, Wait state};

\node[below=0.7 of 14] (N) {};

\draw[arr] (1) edge  (2);
\draw[arr] (1) edge  (3);
\draw[arr] (2) edge node[left]{\tiny 1, clock\_Advance(0)} (4);
\draw[arr] (4) edge  (5);
\draw[arr] (4) edge  (6);
\draw[arr] (5) edge node[left]{\tiny 1, clock\_Advance(1)} (7);
\draw[arr] (7) edge  (8);
\draw[arr] (8) to[out=0, in=180]  node[below, xshift=5]{\tiny 1, init\_transition(2)} (9);
\draw[arr] (3) to[out=-90,in=0]   node[right]{\tiny 1, init\_transition(0)}(9);
\draw[arr] (6) edge  node[left]{\tiny 1, init\_transition(1)} (9.north);
\draw[arr] (9) edge  (10);
\draw[arr] (10) edge node[left]{\tiny 0.01, send\_transition\_fail(0)} (11);
\draw[arr] (10) edge node[right]{\tiny 0.99, send\_transition\_success(0)}  (12);
\draw[arr] (11) edge  (13);
\draw[arr] (13) edge  node[left]{\tiny 1, clock\_Advance(0)}(14);

\draw [arrr] (14) -- (N);
\end{tikzpicture}}
\caption{
Resulting MDP (partial) for the probabilistic timed automata shown in \Cref{PTAExample}.
}
    \label{example2AgentTransitions}   
\end{figure}
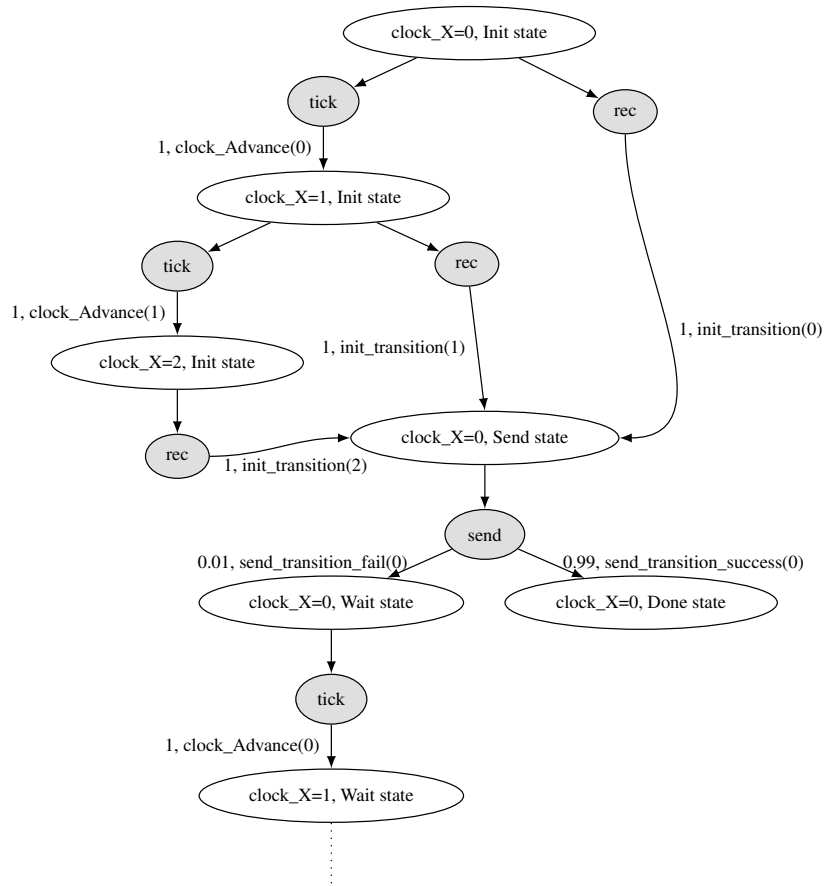

\subsection{PTA example}\label{TimedAutomataExample}
We apply our approach to the probabilistic timed automata example shown in \Cref{PTAExample} and recreated from~\cite{PTAdigitalClock}. 
This simple communication system attempts to send data based on the constraints over clock $x$. 
Here we do not model the actual sends, only the state machine the system goes though.
This is somewhat unnatural for bigraphs, where we are more concerned with global system updates (possibly of multiple agents) than encoding a particular state machine for an entity, but is used to illustrate that we can recover existing PTA semantics.

The system starts in the initial state $\mathtt{Init}$. When the time has elapsed exactly $2$ time units (as forced by the transition constraint \emph{and} the clock invariant on the state) the system moves to the $\mathtt{Send}$ state. The clock invariant $x \le 0$ forces data to be sent immediately. With probability $0.99$ the system reaches its final state, $\mathtt{Done}$, and with $0.01$ probability it fails and moves to a  $\mathtt{Wait}$ state. The system waits at least $4$ time units and at most $8$ time units (forced by the invaraints) before moving back to the $\mathtt{Send}$ state to retry. The clock is reset with each transition.

We start modelling this PTA example by defining the required time constraints.
Here we use $\ctrl{X}(n)$ instead of $\ctrl{LC}(n)$ to be closer to the original example.
For each state we define the valid clock valuations as follows. $\rr{init\_transition}$ rule (\Cref{initTran}) applies over $n \in \{0,1,2\}$ for $\mathtt{Init}$ state, and the rules that are associated to $\mathtt{Send}$ state (\Cref{sendTranSucc} and \Cref{sendTranFail}) fire immediately upon receiving data \ie at $n=0$. While $\rr{wait\_transition}$ rule (\Cref{waitTran}) is applicable over $n \in \{4, 5, 6, 7, 8\}$. 
For all parameters except $2$ for $\mathtt{Init}$ and $8$ for $\mathtt{Wait}$ states, the rules application should be in the same priority class as $\rr{clock\_advance}(\dots)$ rule allowing the non-deterministic behaviour: time passes or an data sends. However, as the invariants force leaving if more than $\maxT{}$ units elapse, $\rr{init\_transition}(2)$ $\rr{send\_transition}(0)$, and $\rr{wait\_transition}(8)$ are in a higher priority class. We encode a single rule for each system transition, and show how the probabilities are encoded using rule weights\footnote{In this case, as there is only one agent moving state, the weights are equivalent to their probabilities since there will never be additional rule application matches to account for.}, \eg to allow $\mathtt{Send}$ state to move to either $\mathtt{Wait}$ or $\mathtt{Done}$.

In \Cref{example2AgentTransitions}, we show the clocks bigraph model conforms to the probabilistic timed automata example by giving the transition system. For space, we only provide the first few transitions. Here the system moves from the initial state $\mathtt{Init}$ to the $\mathtt{Send}$ state. When the system is in the $\mathtt{Init}$ state and the clock constraints are satisfied, there are two possibilities: the time passes or an action occurs. Once the clock becomes $\ctrl{X}(2)$, the system \textbf{has to} move to $\mathtt{Send}$.

\subsection{Cloud System}\label{cloud}
We remodel the example presented in~\cite{timedBig}. The authors also add time constructs to BRSs, but their work differs as it expresses clocks as nested entities and needs multiple reaction rules to advance clocks (which can cause unnecessary state-space explosion).
They do not have a method to track approach wall-time and, as they do not use action BRSs, cannot model the (explicit action) non-deterministic behaviour of many real-time systems.

\begin{figure}[!htb]
\centering
\begin{subfigure}[b]{0.8\textwidth}
\centering
%\resizebox{0.8\textwidth}{!}{\input{tikztext/cloudSendingRule.tex}}
\resizebox{0.8\textwidth}{!}{\includegraphics[width=\linewidth]{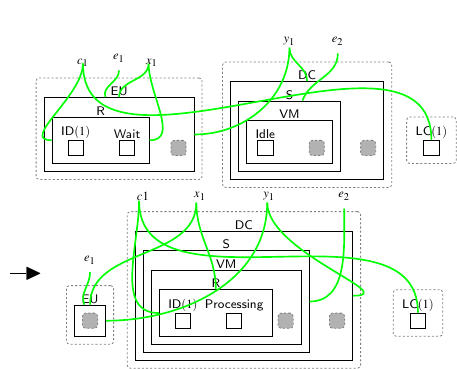}}
\caption{Reaction rule $\rr{sendingRequest}$.
}\label{CloudSendingRuleExample}
\end{subfigure}

\begin{subfigure}[!htb]{0.8\textwidth}
%\resizebox{0.8\textwidth}{!}{\input{tikztext/cloudRetuenRules.tex}}
\resizebox{0.8\textwidth}{!}{\includegraphics[width=\linewidth]{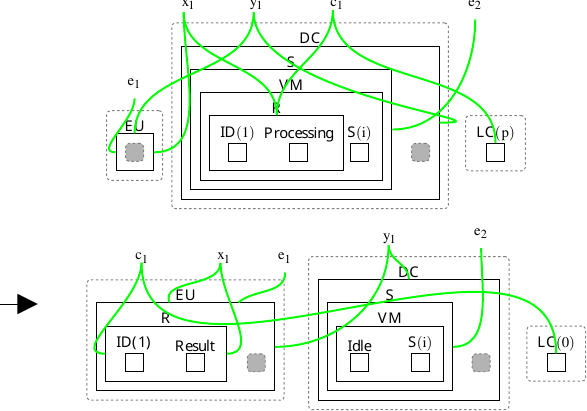}}
\caption{Reaction rule $\rr{returnRequest}(p,i)$, $p=3$ when $i=1$ and $p=4$ when $i=2$.}\label{CloudReturnRuleExample}
\end{subfigure}
\caption{Processing rules for the first request. }
\label{cloudSystemRules}
\end{figure}

A cloud system is a collection of resources that include hardware, software, networks, \etc that is used to store, manage, and process data. That is end users (\ctrl{EU}) send their requests (\ctrl{R}) over the internet to virtual machines (\ctrl{VM}) hosted on physical servers ($\ctrl{S}(i)$ with $i > 0$) within a data centre (\ctrl{DC}) for processing. Standard bigraphs can model and analysis the structure and dynamic behaviour of cloud systems but not the time constraints that cloud applications usually have \eg the duration of tasks. 
We consider a system that has two end users connected to each other via a link \eg $e_1$; and two servers that are also connected via a link \eg $e_2$. A data centre and an end user relate to each other via a link \eg $y_1$. There are four requests each of them connects to its end user through a link \eg $x_1$. The requests need to be processed using two servers. We model the same example as follows. We give each request a parameterised identifier ($\ctrl{ID}(r)$, where $r\:\in\:{1, 2, 3, 4}$). We also associate each request with different controls to reflect its current status. Initially all requests are in a (\ctrl{Wait}) status; requests that are under processing have status \ctrl{Processing} while status \ctrl{Result} indicates that a request has been processed and returned to the end user.
We define a clock for each request but, unlike in~\cite{timedBig}, we add  a global clock that shows the wall-time.
We also use a reaction rule that simultaneously advances all the clocks.
We specify one reaction rule $\rr{sendingRequest}$ (\Cref{CloudSendingRuleExample}) to send a request from an end user to a virtual machine for processing. Given that the servers have varying resources and thus require different amounts of time to process a request, two rules are applied to return the requests after processing, \ie once a predefined time has elapsed. Rule $\rr{processRequest\_S1}$ is for server $\ctrl{S}(1)$ that takes $2$ time units and rule $\rr{processRequest\_S2}$ is for server $\ctrl{S}(2)$ that needs $3$ time units to process a request.
We define the same time constraint assumptions stated in~\cite{timedBig} as integer sets. These sets are to be utilised with the timed reaction rules. That is, for each request there are four integer sets as follows. The \emph{first one} is the valid clock valuations for a request. The \emph{second} contains the time at which a request can be sent. The \emph{last one} is two different sets that determine the time that a request should last according to the processing time for each server. We provide the four integer sets as follows:
\begin{itemize}
    \item $N = \{n \mid 0\leq n \leq \maxT{}\}$ for each request,
    \item $S =\{1,3,4,5\}$ for requests 1, 2, 3, and 4 respectively,
    \item $P_1 = \{s+2 \mid s \in S\}$ for requests processed on server $\ctrl{S}(1)$,
    \item $P_2 = \{s+3 \mid s \in S\}$ for requests processed on server $\ctrl{S}(2)$.
\end{itemize}

While in~\cite{timedBig} virtual machine migrations are considered, here we allocate the process to a machine that is in \ctrl{Idle} status indicating it is free and ready to receive a request. We model sending requests through reaction rule $\rr{sendingRequest}$ (\Cref{CloudSendingRuleExample}). This rule allocates the request to an \ctrl{Idle} server when a request's time constraint is satisfied. We then encode two rules that return requests back to the sender.
Note that depending on the allocated server, only one of the return rules applies: rule $\rr{returnRequest\_S1}$ takes the result back to \ctrl{EU} for requests that are treated on server $\ctrl{S}(1)$ once $2$ time units elapsed; while rule $\rr{returnRequest\_S2}$ takes it back to \ctrl{EU} after $3$ time units passes for requests that are treated on server $\ctrl{S}(2)$. \Cref{CloudReturnRuleExample} shows a return rule that applies when the time constraint (either $p_1$ or $p_2$) is met for $\ctrl{S}(i)$. We will show in the following section how our approach can be used to verify the interesting properties. 
We make the assumption that a request is always forced to send no more than 1 time unit \emph{after} the deadline. This means processing time for delayed tasks can be faster than expected, \eg $p_1 = s + 2 - 1$. This could be accounted for using additional entities, \eg \ctrl{Delayed}, or out of time requests could be dropped. Bigraphs give the flexibility to experiment with these approaches with additional rules.

\subsection{Model Analysis}
To check the feasibility of our proposed modelling technique, we export the bigraph models into MDPs automatically using BigraphER. We can then verify them against required properties using PRISM by expressing properties in the Probabilistic Temporal Logic (PCTL)~\cite{PCTL1}. 
To help writing properties, we utilise \emph{bigraph patterns} that allow states to be labelled based on matches~\cite{savannah}, \eg label any state that has a \ctrl{Data} entity.
Since clocks are just part of the model, we can also label clock values, \eg $clock\_LC_0$ for $\ctrl{LC(0)}$.
Interestingly, the clocks used in the properties are those in the model, \ie they are just bigraph entities, rather than clock variables you would generate if you, for example, expressed the PTA directly in PRISM.
This gives flexibility, \ie we can write predicates over many different types of clock matches. 

We make use of the following PCTL quantifiers to specify our properties: $\mathbf{A}$ (for all) and $\mathbf{E}$ (there exists), and path formulae, $\mathbf{F}$ (eventually), $\mathbf{X}$ (next), $\mathbf{U}$ (until) and $\mathbf{\wedge}$ (and). We also use probability quantifier $\mathbf{P}$ to check the probability of reaching a specific state. 
For example, $\mathbf{P} \leq 0.5 \left[\, \mathbf{F}\, state A \,\right]$ means: (over all paths) is there at most a $0.5$ probability that the system eventually reaches a state satisfying $state A$, we label a state as $state A$ using a bigraph pattern. 

We conduct model checking on several interesting properties. In the PTA example (\Cref{TimedAutomataExample}), we can perform probabilistic reachability properties such as ``Is there at least a $0.99$ percent probability that the system moves to $\mathtt{Done}$ state successfully'', which holds for this system.

\begin{equation}
    \mathbf{P} \geq 0.99\:\left[\:\mathbf{F} \:\left(\mathtt{In\_Done\_state}\right)\:\right]
\end{equation}
We can also utilise clock predicates to check properties relating to time \eg ``eventually there is at least a $99\%$ chance that the system moves to the $\mathtt{Done}$ state and resets clock X'', which is true for our system.
Importantly, these are based on our clock entities, not time-bounded properties within PRISM which operate on their own time units. 

\begin{equation}
    \mathbf{P}\geq 0.99\:[\:\mathbf{F}\:\left(\: \mathtt{In\_Done\_state}\:\mathbf{\wedge}\:\mathtt{clock\_X_0}\right)\:]
\end{equation}
We can check clock invariants are satisfied, \eg that the system is not be in $\mathtt{Wait}$ state after $8$ time units have elapsed:

\begin{equation}
    \mathbf{A}\: [\:\neg \left(\mathbf{F}\:(\mathtt{In\_Wait\_state})\: \mathbf{\wedge}\: \left(\mathtt{clock\_X_9}\right) \: \right)\:]\quad 
    \;
\end{equation}
As expected this is true in all cases.

For the cloud system we can, for example, check the first request is always sent once $one$ time unit has elapsed.
As we use next (\textbf{X}), the request must be sent immediately before any other state transition. 

\begin{equation}
    \mathbf{A} \: [\:\mathbf{F}\: \:\mathtt{clock_1}\: \mathbf{\wedge}\:\left(\mathbf{X}\:\: \mathtt{request_1\_sent\_to\_S}\right)\:] 
\end{equation}

\section{Related work}\label{relatedwork}
Many modelling formalisms have been extended to real-time systems by introducing clocks. Timed CCS~\cite{CCS+time} extends Milner’s CCS by introducing the time notion to create concurrency models for real-time systems. 
Timed CCS introduces a variable to record time delays \eg before a message arrives. 
Similar to our work, timed CCS considers the positive real number including zero as the time domain for convenience, but the model can deal with another numerical domain \eg the natural numbers.
Timed Petri nets \cite{PetriNets} extends Petri nets to allow time triggered transitions. It uses tokens to allow transitions to start and then they are placed into the output when the firing process terminates. 
The authors use random variables with continuous or discrete probability distribution functions for the firing time. Timed $\pi$-Calculus~\cite{TimedCalculus} extends the $\pi$-Calculus with continuous time to describe and reason about concurrent Cyber-Physical Systems and real-time systems. An executable operational semantics of $\pi$-Calculus is developed in Logic Programming to model concurrency. 

Graph Transformation Systems (GTS) has been extended into Probabilistic Timed Graph Transformation Systems (PTGTSs) which enable modelling and analysing structure dynamic and timed and probabilistic behaviour of embedded systems~\cite{GTS}. The clock is a typed node that is contained in a graph to identify the nodes that are used for time measurement only. In this work, PTGTSs is formally mapped into Probabilistic Timed Automata (PTA) where the Probabilistic Timed Structure (PTS) of the mapped PTGTSs is equal to the PTS of the resulting PTA, hence they both satisfy the same set of PTCTL properties. The obtained PTA can be checked using PRISM. However, the mapping process does not consider three aspects of the PTA. First, since the PTA does not consider valuations in the labelling function, constraints are ignored in the mapping process so the constraint should be true for any such atomic proposition. It also considers the PTGTS that does not show timelocks during its execution instead of finding and removing them when mapping PTGTS into PTA. Finally, the GTS state space that is constructed for PTGTS is up to isomorphism as preserved clock nodes are respected which may affect the state space finiteness of the resulting PTA. In our work, we bound the clock valuations which ensures finite transition systems. Additionally, in case that a system frequently resets its local clocks, the employment of the global clock guarantees the finiteness of the transition system. To prevent timelocks and so obtain a proper transition between reachable states, we assign the maximum states invariant valuation to the rules that are in a higher priority.  

Bigraphs have been applied to model the location and connectivity of components of structure-aware mobile systems using BigrTimo that combines the rTiMO process algebra and bigraphs~\cite{mobileSystems}. It uses real-time constraints to control actions by showing the waiting time for communication. The work in~\cite{timedBig} explicitly encodes clocks as entities within bigraphs to model and reason about cloud applications, and shares some similarities with our approach as discussed in \cref{cloud}.
It adds a set of clocks and a set of clocks constraints that are associated with nodes. It then utilises two different types of rules:  (1) a set of reaction rules to advance all clocks, all clocks are advanced at the same speed; (2) instantaneous rules\footnote{Silent rules that do not appear in an output transition system.} that are executed only when there is a match and time constraints of one or many nodes are satisfied. 
These rules can also update the clock constraints and resets the clocks.
However, when a new time constraint is satisfied, the time cannot elapse and another instantaneous rule may apply subsequently. In contrast, we encode timed aspects as action bigraphs resulting in an MDP transition system that explicitly models the non-deterministic behaviour of timed systems, that then can be formally checked by different model checking tools.
The work uses Real-Time Maude language~\cite{maude} and its TCTL model checker to implement and analyse the approach. 
State-space explosion is reduced here by employing a strategy that models all clocks as a separate region allowing us to use only one reaction rule to advance all clocks simultaneously. We imitate the wall-time by utilising the global clock entity. 

\section{Conclusion}\label{conclusion}
Bigraphical Reactive Systems (BRSs) have been successfully used to model a wide range of systems but, until now, they had no explicit support for real-time systems. 
We overcome this limitation by introducing a modelling technique that uses the digital clocks approximation of (probabilistic) timed automata to encode timing aspects. 
This approach relies on Action Bigraphs, which have as a semantics Markov Decision Processes (MDPs), meaning we can express both probabilistic and timing behaviour within models.
Using BigraphER, we encode the proposed strategy and verify the MDP corresponding to each model using PCTL model checking.
Using two examples, we show our approach supports multiple clocks and the transition system obtained via rewriting faithfully encodes the digital-clock approximation of the behaviour of a real-time system.

Our approach suffers from the same limitations as the digital-clock approximation \ie currently we do not support diagonal clocks and strict inequalities.
We mitigate state space explosion by adopting the following two strategies.
First, we bound clock valuations thus we obtain finite transition systems. 
Second, we allow the base time unit to be specified in each model, effectively allowing clocks to advance by multiple ticks in one step.
Another limitation of our approach is that we assume clocks are always synchronised, \ie they all progress at the same speed, which might not always be the case in scenarios like wireless sensor networks and IoT.

In future, we will develop syntactic support for clock constraints in the BigraphER language to generate real-time ABRS models like the ones considered in the paper. 
Sorting schemes~\cite{sorts} (type systems for bigraphs) will ensure correct separation of clocks and model entities.
Syntax and sorts will ensure our extended set of reaction rules is correct by construction. 
We also aim to extend our approach to also support diagonal clocks and strict inequalities.

\section*{Acknowledgement}
This work is supported by the UK EPSRC projects CHEDDAR (EP/X040518/1) and CHEDDAR Uplift (EP/Y037421/1), and an Amazon Research Award on Automated Reasoning.

%\newpage
\bibliographystyle{eptcs}
\bibliography{Bibliography1}

\newpage

\begin{appendices}
\section{BigraphER implementation of the examples in Section 4}
\begin{bigrapher}
#################  PTA Example #################

atomic fun ctrl X(n) = 1 ;
ctrl S = 1;
atomic ctrl Init = 0;
atomic ctrl Send = 0;
atomic ctrl Wait = 0;
atomic ctrl Done = 0;

#################  Reaction Rules #################

##Init
fun react init_transition(n) =
  S{c}.Init || X(n){c}
  -[1]->
  S{c}.Send || X(0){c};

## Send
fun react send_transition_success(n) =
  S{c}.Send || X(n){c}
  -[0.99]->
  S{c}.Done || X(n){c};
  
fun react send_transition_fail(n) =
  S{c}.Send || X(n){c}
  -[0.01]->
  S{c}.Wait || X(n){c};


# Wait
fun react wait_transition(n) =
  S{c}.Wait || X(n){c}
  -[1]->
  S{c}.Send || X(0){c};

#Done
react done_done =
  S{c}.Done -[1]-> S{c}.Done;
  
################# Clock Advance Rule #################
fun react clock_advance(n) =
  X(n){c}
  -[1]->
  X(n + 1){c};

################# Predicates #################
fun big clock_X(n) = X(n){c};
big in_Init_state = S{c}.Init;
big in_Send_state = S{c}.Send;
big in_Wait_state = S{c}.Wait;
big in_Done_state = S{c}.Done;

################# Initial State #################
big example_PTA = /c (S{c}.Init || X(0){c});

 begin abrs
  int n = {0,1,2,3,4,5,6,7,8};
  int maxInitT = {2};
  int init_Sending_Time = {0,1};
  int maxSendT = 0;
  int maxWaitT = 8;
  int wait_Sending_Time={4,5,6,7};
  
  init example_PTA;

  rules = [
    # Higher in the list => higher priority
    {done_done, 
     init_transition(maxInitT), 
     send_transition_fail(maxSendT),  
     send_transition_success(maxSendT), 
     wait_transition(maxWaitT)},
     
    {clock_advance(n), 
     wait_transition(wait_Sending_Time),
     init_transition(init_Sending_Time)}
  ];

  actions = [
    send={send_transition_success,send_transition_fail},
    retry={wait_transition},
    rec={init_transition},
    deadlock={done_done},
    tick={clock_advance}
  ];
  
  preds = {
    in_Init_state,
    in_Send_state,
    in_Wait_state,
    in_Done_state,
    clock_X(n)
  };  
end
\end{bigrapher}

 \newpage

\begin{bigrapher}
#################  Cloud System Example #################

ctrl FrontEnd = 0;
ctrl EU = 3;
ctrl R = 2;
atomic ctrl Processing = 0;
atomic ctrl Result = 0;
atomic ctrl Wait = 0; 
atomic ctrl Idle = 0 ;
atomic ctrl Stop = 0 ;
ctrl BackEnd = 0;
ctrl DC = 1;
ctrl S = 1;
atomic ctrl S1 = 0;
atomic ctrl S2 = 0;
ctrl VM = 0;
atomic fun ctrl ID(i) = 0;
ctrl LocalClock = 0;
atomic fun ctrl LC(requestClock) = 1;
atomic fun ctrl GC(globalClock) = 0;

################# Reaction Rules #################
fun react clock_advance(request1Clock, request2Clock, request3Clock, request4Clock, gc) = 
  LocalClock.( LC(request1Clock){loclock1} | LC(request2Clock){loclock2} | LC(request3Clock){loclock3} | LC(request4Clock){loclock4} ) | GC(gc) 
  -[1]-> 
  LocalClock.( LC(request1Clock +1 ){loclock1} | LC(request2Clock +1 ){loclock2} | LC(request3Clock +1 ){loclock3} | LC(request4Clock +1 ){loclock4} ) | GC(gc+1) if ! Stop in ctx; 

fun react sendingRequest(i, r1_Send_Time) =     
  EU{x1,y1,e1}.(R{x1,c1}.(ID(i) | Wait ) | id ) ||  DC{y1}.( S{e2}.VM.(Idle | id ) | id )  || LC(r1_Send_Time){c1} 
  -[1]-> 
  EU{x1,y1,e1}.id  ||  DC{y1}.( S{e2}.VM.(R{x1,c1}.(ID(i)  | Processing) | id ) | id  )  || LC(r1_Send_Time){c1} ;

fun react returnRequest_S1(i, s1_Process_Time) = 
  EU{x1,y1,e1}.id  ||  DC{y1}.( S{e2}.VM.(R{x1,c1}.(ID(i)  | Processing) | S1 ) | id  )  || LC(s1_Process_Time){c1}      
  -[1]-> 
  EU{x1,y1,e1}.( R{x1,c1}.( ID(i) | Result ) | id ) ||  DC{y1}.( S{e2}.VM.(Idle | S1 ) | id ) || LC(0){c1};

fun react returnRequest_S2(i, s2_Process_Time) = 
  EU{x1,y1,e1}.id ||  DC{y1}.( S{e2}.VM.(R{x1,c1}.(ID(i)  | Processing) | S2 ) | id  ) || LC(s2_Process_Time){c1}      
  -[1]-> 
  EU{x1,y1,e1}.( R{x1,c1}.( ID(i) | Result ) | id ) ||  DC{y1}.( S{e2}.VM.(Idle | S2 ) | id )  || LC(0){c1};

react done =  
  EU{x1,x2,e1}.( R{x1,c1}.(Result | id ) | R{x2,c2}.(Result | id ) ) | EU{y1,y2,e1}.( R{y1,c3}.( Result | id ) | R{y2,c4}.(Result | id ) )
  -[1]-> 
  EU{x1,x2,e1}.( R{x1,c1}.(Stop | id ) | R{x2,c2}.(Stop | id ) ) | EU{y1,y2,e1}.( R{y1,c3}.( Stop | id ) | R{y2,c4}.(Stop | id ) ) ; 


################# Initial State #################
big cloudSystem = 
/x1/x2/y1/y2/e1/e2/c1/c2/c3/c4 (
 FrontEnd.( 
   EU{x1,x2,e1}.( R{x1,c1}.(Wait | ID(1) ) | R{x2,c2}.(Wait | ID(2) ) ) | EU{y1,y2,e1}.( R{y1,c3}.( Wait | ID(3) ) | R{y2,c4}.(Wait | ID(4) ) ) 
)
|| BackEnd.(DC{e1}.( S{e2}.VM.(Idle | S1 ) | S{e2}.VM.(Idle | S2 )) )
|| LocalClock.( LC(0){c1} | LC(0){c2} |  LC(0){c3} | LC(0){c4} ) | GC(0)
);

################# Predicates #################
fun big request_Sent_to_S1_at(i, x) = 
   S{e2}.VM.(R{x1,c1}.(ID(i) | Processing ) | S1 )  
|| LC(x){c1};

fun big request_Return_at(i, x) = 
R{x1,c1}.(ID(i) | Result ) || LC(x){c1};


begin abrs
  int i = {1, 2, 3, 4};
  int request1Clock={0,1,2,3,4,5,6,7,8};
  int request2Clock={0,1,2,3,4,5,6,7,8};
  int request3Clock={0,1,2,3,4,5,6,7,8};
  int request4Clock={0,1,2,3,4,5,6,7,8};
  int gc={0,1,2,3,4,5,6,7,8,9,10,11,12};
  
  int r1_Send_Time={1};
  int r1_Max_Send_Time={2};
  int r2_Send_Time={3};
  int r2_Max_Send_Time={4};
  int r3_Send_Time={4};
  int r3_Max_Send_Time={5};
  int r4_Send_Time={5};
  int r4_Max_Send_Time={6};
  
  int r1_Process_Time_S1={3};
  int r1_Process_Time_S2={4};
  int r2_Process_Time_S1={5};
  int r2_Process_Time_S2={6};
  int r3_Process_Time_S1={6};
  int r3_Process_Time_S2={7};
  int r4_Process_Time_S1={7};
  int r4_Process_Time_S2={8};

  int x ={0,1,2,3,4,5,6,7,8,9,10,11};


init cloudSystem;

  rules = [ 
        {done},
        {returnRequest_S1(1, r1_Process_Time_S1), returnRequest_S1(2, r2_Process_Time_S1), returnRequest_S1(3, r3_Process_Time_S1), returnRequest_S1(4, r4_Process_Time_S1),  

        returnRequest_S2(1, r1_Process_Time_S2), returnRequest_S2(2, r2_Process_Time_S2), returnRequest_S2(3, r3_Process_Time_S2), returnRequest_S2(4, r4_Process_Time_S2)},

        {sendingRequest(1,r1_Max_Send_Time), sendingRequest(2,r2_Max_Send_Time), sendingRequest(3,r3_Max_Send_Time), sendingRequest(4, r4_Max_Send_Time)} ,

        {sendingRequest(1,r1_Send_Time), sendingRequest(2,r2_Send_Time), sendingRequest(3,r3_Send_Time), 
        sendingRequest(4, r4_Send_Time), clock_advance(request1Clock,request2Clock, request3Clock, request4Clock, gc)}
   ];

actions=[
 send={sendingRequest},
 return = { returnRequest_S1, returnRequest_S2 },
 tick = {clock_advance},
 stop = {done}
];

preds = {
request_Sent_to_S1_at(i, x), request_Return_at(i, x)};
end
\end{bigrapher}

\end{appendices}
\end{document}